\documentclass{PoS}

\usepackage{graphicx,times}
\usepackage{epsf}
\usepackage{amsmath,amssymb,latexsym,float}
\usepackage{enumerate}
\usepackage{listings}
\newtheorem{thm}{Theorem}[section]

\numberwithin{equation}{section}
\newcommand{\norm}[1]{\left\Vert#1\right\Vert}

\newcommand{\R}{\text{\fontshape{n}\selectfont I\kern-.42exR}}

\newcommand{\1}{\text{\fontshape{n}\selectfont 1\kern-.56exl}}

\newcommand{\qed}{\hspace*{\fill}\text{$\Box$}}

\title{A Schur Complement Approach to Chiral Fermions}

\ShortTitle{A Schur Complement Approach to Chiral Fermions}

\author{\speaker{Artan Bori\c{c}i}\\
        University of Tirana\\
        Department of Physics, Faculty of Natural Sciences\\
        King Zog I Boulevard, Tirana, Albania\\
        E-mail: \email{borici@fshn.edu.al}}

\abstract{
\vspace{1cm}
Lattice chiral fermions are synonymous to the Ginsparg-Wilson relation \cite{GW}. Indeed, this relation is satisfied by the overlap \cite{NaNe93,Ne98}, domain wall \cite{Ka92,FuSha95} and perfect action \cite{Hasenfr_et_al98} fermion kernel.

\vspace{0.15cm}
In a recent work we have shown that it is possible to take a direct RG approach for fermions in the presence of gauge fields \cite{Borici07}. This is due to an algebraically implicit blocking technique which yields a Schur-complementary coarse Dirac operator. Using a Schur complement approximation which is stable and regular, the scheme can be iterated to the fixed point.

\vspace{0.15cm}
In this talk, we elaborate more on the direct RG approach and show how to get highly improved chiral fermions on the coarse lattice with the gauge fields remaining on the fine lattice. We give numerical examples in the case of lattice QCD using QCDLAB {\tt http://phys.fshn.edu.al/qcdlab.html}.

}

\FullConference{The XXV International Symposium on Lattice Field Theory\\
		 July 30-4 August 2007\\
		 Regensburg, Germany}

\begin{document}

\section{Renormalisation group transformations as partial Gaussian integrals}

A renormalisation group (RG) transformation for quadratic actions, as it is the case of lattice fermions, is a simple Gaussian integration,
\begin{equation}\label{rgt}
\det\tilde D~e^{-\bar\psi_b S_{bb} \psi_b}=\int_{\bar\phi\phi}e^{-(\bar\psi_b-\bar\phi\bar B)D_{bb}(\psi_b-B\phi)-\bar\phi D\phi}\ ,
\end{equation}
where $D$ and $S_{bb}$ are Dirac operators on the fine and coarse lattices, $D_{bb}$ is a coarse lattice kernel, $B,\bar B$ are blocking operators, and, by evaluating the right hand side, one can show that
\begin{equation}\label{rg}
\tilde D=D+\bar B D_{bb}B, ~~~~~~~S_{bb}=D_{bb}-D_{bb}B{\tilde D}^{-1}\bar BD_{bb}\ .
\end{equation}

The difficulty with fermion RG transformations in lattice gauge theories is related to the special properties of the blocking kernels $B$ and $\bar B$. They should be such that the coarse operator, $S_{bb}$, and its approximation, $\tilde S_{bb}$, satisfy:
\begin{itemize}
\item[1.] {\it gauge covariance};
\item[2.] {\it stability}
\footnote{The notions of stability and regularity used here do not necessarily correspond to those of the mathematical literature.}
, i.e. inheritance of the fine lattice operator properties;
\item[3.] {\it regularity}, i.e. $\norm{I_b-\tilde S_{bb}^{-1}S_{bb}} < 1$, or inheritance of the infrared mode physics.
\end{itemize}
These properties are important ingredients of a valid RG scheme that can be iterated to the fixed point. Indeed, non-covariant blocking prescriptions fail to preserve the covariance of the original operator. Stability ensures the conservation of algebraic properties and the locality of the fermion theory, whereas regularity is essential for the preservation of the low lying modes of the Dirac operator.

Finding such kernels by direct search is a daunting task. For free fermions, this task can be simplified a lot as is shown by the authors of classically perfect actions \cite{Hasenfr_et_al98}.

Our approach simply requires that the coarse lattice operator is the Schur complement of a block UL-decomposition of the fine lattice operator \cite{Borici07}. In this way, gauge covariance of the coarse operator is guaranteed, whereas $B,\bar B$ are derived rather then defined. If we partition the Dirac operator as a 2x2 block operator,
\begin{equation}\label{2x2}
D=
\begin{pmatrix}
D_{bb} & D_{br}\\
D_{rb} & D_{rr}
\end{pmatrix}\ ,
\end{equation}
its block UL decomposition yields,
\begin{equation}\label{ul}
D=
\begin{pmatrix}
I_{bb} & D_{br}D_{rr}^{-1}\\
0 & I_{rr}
\end{pmatrix}
\begin{pmatrix}
S_{bb} & 0\\
D_{rb} & D_{rr}
\end{pmatrix},~~~~~~S_{bb}=D_{bb}-D_{br}D_{rr}^{-1}D_{rb}\ ,
\end{equation}
$S_{bb}$ being the Schur complement. In addition, this approach allows us to formulate similar stability and regularity properties as in other applications \cite{reusken00}.

But how can we bridge between a block UL-decomposition and an RG transformation? This is easily done if we notice that the Schur complement appears in the left hand side of the partial Gaussian integration with respect to the $r$-components of the fermion field, i.e.
\begin{equation}\label{pgi}
\det D_{rr}~e^{-\bar\phi_b S_{bb} \phi_b}=\int_{\bar\phi_r\phi_r}e^{-\bar\phi D\phi}=
\int_{\bar\phi_r\phi_r}e^{-\bar\phi_b D_{bb}\phi_b-\bar\phi_b D_{br}\phi_r-\bar\phi_r D_{rb}\phi_b-\bar\phi_r D_{rr}\phi_r}\ .
\end{equation}
Therefore, there is no need to refer to the blocking kernels any more. However, in order to complete the formal argument, it is straightforward to show that using covariant blocking kernels,
$$
B=
\begin{pmatrix}
0, & D_{bb}^{-1}D_{br}
\end{pmatrix}, ~~~~~\bar B=
\begin{pmatrix}
0\\
D_{rb}D_{bb}^{-1}
\end{pmatrix}\ ,
$$
and substituting in (\ref{rg}) one gets the Schur complement as given in (\ref{ul}), whereas comparing equations (\ref{rgt}) and (\ref{pgi}), one identifies $\psi_b\equiv\phi_b$.

\section{Permutation of lattice sites}

For the Schur complement to be the coarse Dirac operator, the $b$-components of the fermion field should be defined on the coarse lattice. Hence, starting form a given ordering, we have to perform a permutation of lattice sites, such that the block sites are labelled first.

We use two permutation types, which yield favourable stability and regularity properties of the Schur complement. For ease of illustration, we will define them in the case of a 2-dimensional lattice (see Figure 1):

\vspace{5.8cm}
{\hspace{-4.5cm}\epsfxsize=8cm \epsffile[10 20 300 180]{lattice.ps}}
\vspace{0.5cm}

{{\bf Figure 1.}\footnotesize ~~Example of a site permutation on a 4x4 lattice. Coarser lattice sites are labelled by boldface font.}

\vspace{0.5cm}
\begin{itemize}
\item {\it Type I permutation.} Split the lattice in $2^2$ blocks and label those in the lower left corner as block lattice sites.
\item {\it Type II permutation.} Split the lattice in even-odd slices along the vertical direction and label even slice lattice sites as block sites. Then, repeat the same for the horizontal direction.
\end{itemize}
The generalisation in four dimensions is straightforward.

For each permutation type we assign matrix operators which perform row permutations of the lattice Dirac operator. Since to every row permutation there is a column permutation, we consider four permutation operators, $P_I,P_{II},P_I^{-1},P_{II}^{-1}$, all applied to the left of $D$.

\section{Schur complement approximation}

It is clear that the Schur complement itself is a full matrix, and hence, not practical for the iteration of the RG scheme. Therefore, one has to rely on some approximation. In this section we show how to construct a stable and regular Schur complement approximation for {\it a non-negative} Wilson Dirac operator, i.e.
$$
D=I - M, ~~~~~~~~\norm{M}\leq 1\ .
$$
In order to fix the idea we consider a Type II permutation,
$$
P_{II}DP_{II}^{-1}=
\begin{pmatrix}
I_b-M_{bb} & ~~-M_{br}\\
~~-M_{rb} & I_r-M_{rr}
\end{pmatrix}\ ,
$$
where $b$ labels {\it even} (3-dimensional) lattice slices transverse a given direction, whereas $r$ labels {\it odd} lattice slices transverse the same direction. Hence, the Schur complement will be,
$$
S_{bb}=I_b-M_{bb}-M_{br}\left(I_r-M_{rr}\right)^{-1}M_{rb}\ ,
$$
which is well defined if $\norm{M_{rr}}<1$. It is easy to compute the norm bounds,
$$
\norm{M_{br}}\leq 2\kappa, ~~\norm{M_{bb}}=\norm{M_{rr}}\leq 6\kappa, ~~\norm{M}\leq 8\kappa\ ,
$$
where $\kappa$ is the usual hopping parameter for Wilson fermions.
Since $D$ must be non-negative, i.e. $\norm{M}\leq 1$, the condition
\begin{equation}\label{kapa}
\kappa\leq \frac18
\end{equation}
will be imposed. In this case we get,
\begin{equation}\label{m_norms}
\norm{M_{br}}\leq \frac14, ~~\norm{M_{bb}}=\norm{M_{rr}}\leq \frac34\ .
\end{equation}
Therefore, $\left(I_r-M_{rr}\right)^{-1}$ can be expressed as geometric series in $M_{rr}$ and the Schur complement takes the form,
\begin{equation}\label{sc}
S_{bb}=I_b-\left(M_{bb}+M_{br}~\sum_{l=0}^{\infty}~M_{rr}^{~l}~M_{rb}\right)\ .
\end{equation}
Now, we are in a position to define the order $k$ approximation,
\begin{equation}\label{sc_approx}
{\tilde S_{bb}^{(k)}}=I_b-\left(M_{bb}+M_{br}~\sum_{l=0}^{k-1}~M_{rr}^{~l}~M_{rb}\right), ~~~k=1,2,\ldots\ .
\end{equation}



\begin{thm}
The coarse Wilson Dirac operator defined on the even 3-dimensional slices of a 4-dimensional lattice is stable and regular for any approximation order $k$ and hopping parameter $\kappa \leq 1/8$.
\end{thm}

{\it Proof}. From eqs. (\ref{sc}) and (\ref{sc_approx}) is clear that 
$$
\norm{M_{bb}+M_{br}~\sum_{l=0}^{\infty}~M_{rr}^{~l}~M_{rb}}\leq \frac34+\frac14~\frac1{1-\frac34}~\frac14 = 1
$$
and
\begin{equation}\label{sc_approx_m_norm}
\norm{M_{bb}+M_{br}~\sum_{l=0}^{k-1}~M_{rr}^{~l}~M_{rb}}\leq \frac34+\frac14~\frac{~~~1-\left(\frac34\right)^k}{1-\frac34}~\frac14 = 1-\frac14\left(\frac34\right)^k\ .
\end{equation}
Therefore, if $D$ is a non-negative operator, then $S_{bb}$ and $\tilde S_{bb}^{(k)}$ are also non-negative operators. This proves the stability of the scheme.

For the regularity, consider,
$$
S_{bb}={\tilde S_{bb}^{(k)}}-M_{br}~\sum_{l=k}^{\infty}~M_{rr}^{~l}~M_{rb}
$$
and then
\begin{equation}\label{distance}
I_b-\left({\tilde S_{bb}^{(k)}}\right)^{-1}S_{bb}=\left({\tilde S_{bb}^{(k)}}\right)^{-1}\left(M_{br}~\sum_{l=k}^{\infty}~M_{rr}^{~l}~M_{rb}\right)
\end{equation}
First, consider the norm of the second term in the right hand side:
$$
\norm{M_{br}~\sum_{l=k}^{\infty}~M_{rr}^{~l}~M_{rb}}\leq \frac14~\sum_{l=k}^{\infty}\left(\frac34\right)^k~\frac14=\frac14~\left(\frac34\right)^k\ .
$$
From, (\ref{sc_approx_m_norm}) we can express $\left(\tilde S_{bb}^{(k)}\right)^{-1}$ as a geometric series. Hence, for the first term in the right hand side of (\ref{distance}) we get:
$$
\norm{\left({\tilde S_{bb}^{(k)}}\right)^{-1}}\leq \frac1{\frac14\left(\frac34\right)^k}
$$
Putting results together, we get:
$$
\norm{I_b-\left({\tilde S_{bb}^{(k)}}\right)^{-1}S_{bb}}\leq 1\ .
$$
This proves the regularity and, hence, the theorem. $\qed$

\bigskip
{\bf Remarks}

\begin{itemize}
\item[1.] In fact, the theorem proves the regularity in its weak form. The strong regularity,
$$
\norm{I_b-\left({\tilde S_{bb}^{(k)}}\right)^{-1}S_{bb}} < 1\ ,
$$
is not proven in this setting, but can be anticipated at large orders, $k$.
\item[2.] It is easy to note that the first order approximation preserves the sparsity pattern, an attractive approximation from the computational point of view. However, from the RG point of view, this corresponds to a decimation of the odd 3-dimensional transversal lattices which does not expand the coupling constant space. This is similar to Migdal-Kadanoff approximation, which does not lead to the expected criticality of the 2-dimensional Ising model \cite{Creutz_book}. Therefore, one should use approximation orders $k>1$.
\item[3.] A similar theorem should hold for Type I permutations. However, numerical results favour inverse Type II permutations \cite{Borici07}.
\end{itemize}

The approximation can be improved further if one modifies the last term in the expansion:
\begin{equation}\label{sc_approx_tilde}
{\tilde S_{bb}^{(k)}}=I_b-\left(M_{bb}+M_{br}~\sum_{l=0}^{k-2}~M_{rr}^{~l}~M_{rb}\right)-M_{br}~M_{rr}^{~k-1}~\left(I_r-\tilde M_{rr}\right)^{-1}~M_{rb},
\end{equation}
where $\tilde M_{rr}$ is a diagonal matrix, its entries being the sum of $M_{rr}$ rows. In the multigrid terminology, this makes the approximation {\it consistent} \cite{reusken00}.
A simple effect of this property is the inheritance of the smallest singular value. For free fermions, the original operator is singular for $8\kappa=1$, whereas using (\ref{sc_approx}) at second order, the smallest singular value is $1-(6\kappa+4\kappa^2+24\kappa^3)>0$. However, using (\ref{sc_approx_tilde}) with $M_{rr}=6\kappa I_r$, the smallest singular value is $1-(6\kappa+4\kappa^2+24\kappa^3/(1-6\kappa))=0$.

\section{Numerical examples}

Figure 2 shows the spectrum of the Schur complement for four different permutations. For inverse permutation types and a suitable shift, $c$, the shifted Schur complement can be used as an improved domain-wall/overlap kernel. In Figure 3 we show the spectrum of $\left({\tilde S_{bb}^{(2)}}\right)^{-1}S_{bb}$ using (\ref{sc_approx_tilde}). All permutation types show excellent regularity properties.

In summary, the Schur complement RG approach is a very promising device for lattice QCD computations: we have shown how to get improved chiral kernels; algorithmically, the scheme yields, for the first time, a working multigrid algorithm \cite{Borici07}.

{\hspace{-2cm}\includegraphics[width=17cm,height=11cm]{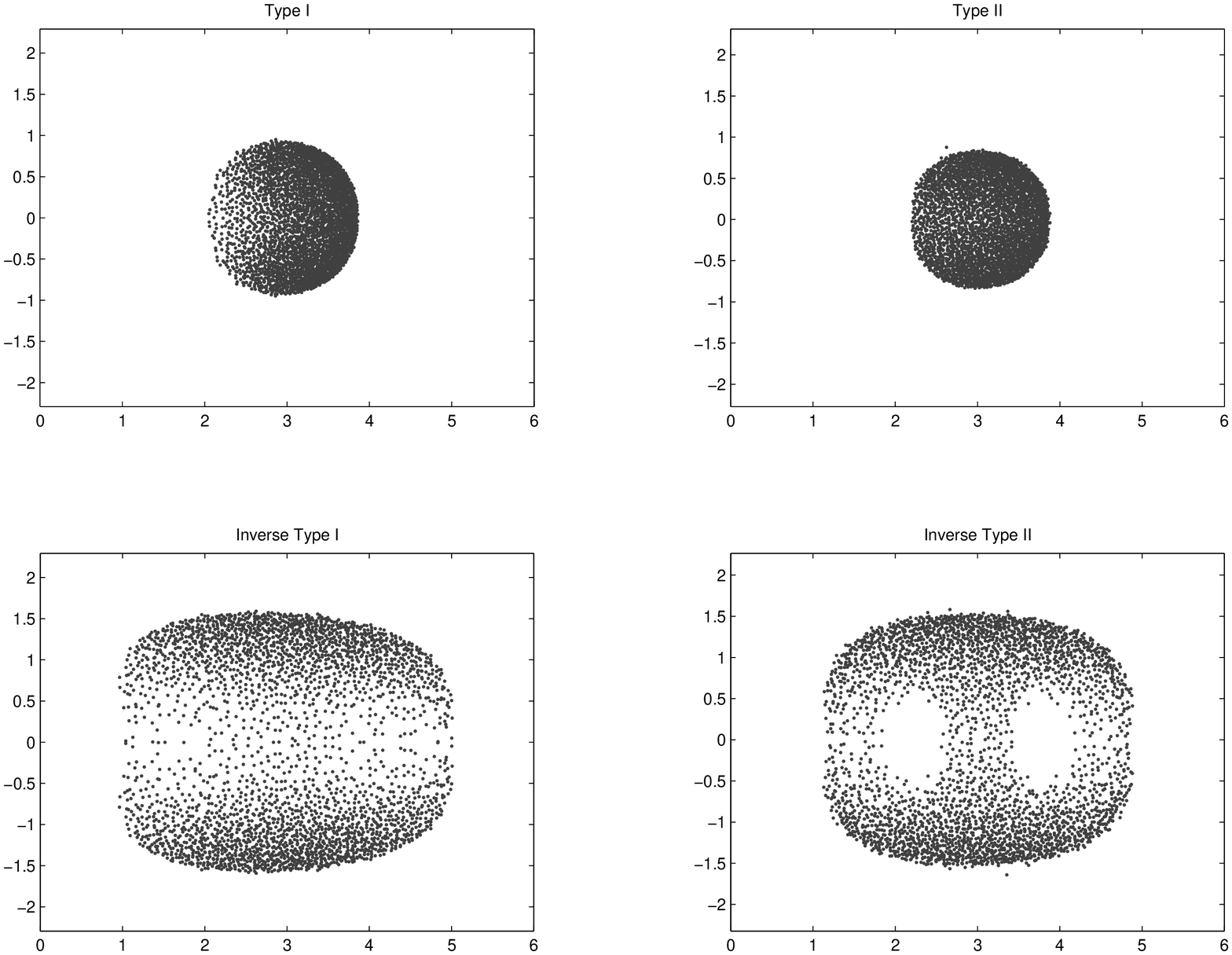}}

\vspace{-0.5cm}
{{\bf Figure 2.}\footnotesize ~~Spectrum of $S_{bb}$ for various permutations in a SU(3) background field at $\beta=5.4$ on a $8^8$ lattice.}

{\vspace{1cm}\hspace{-1.3cm}\includegraphics[width=16cm,height=10cm]{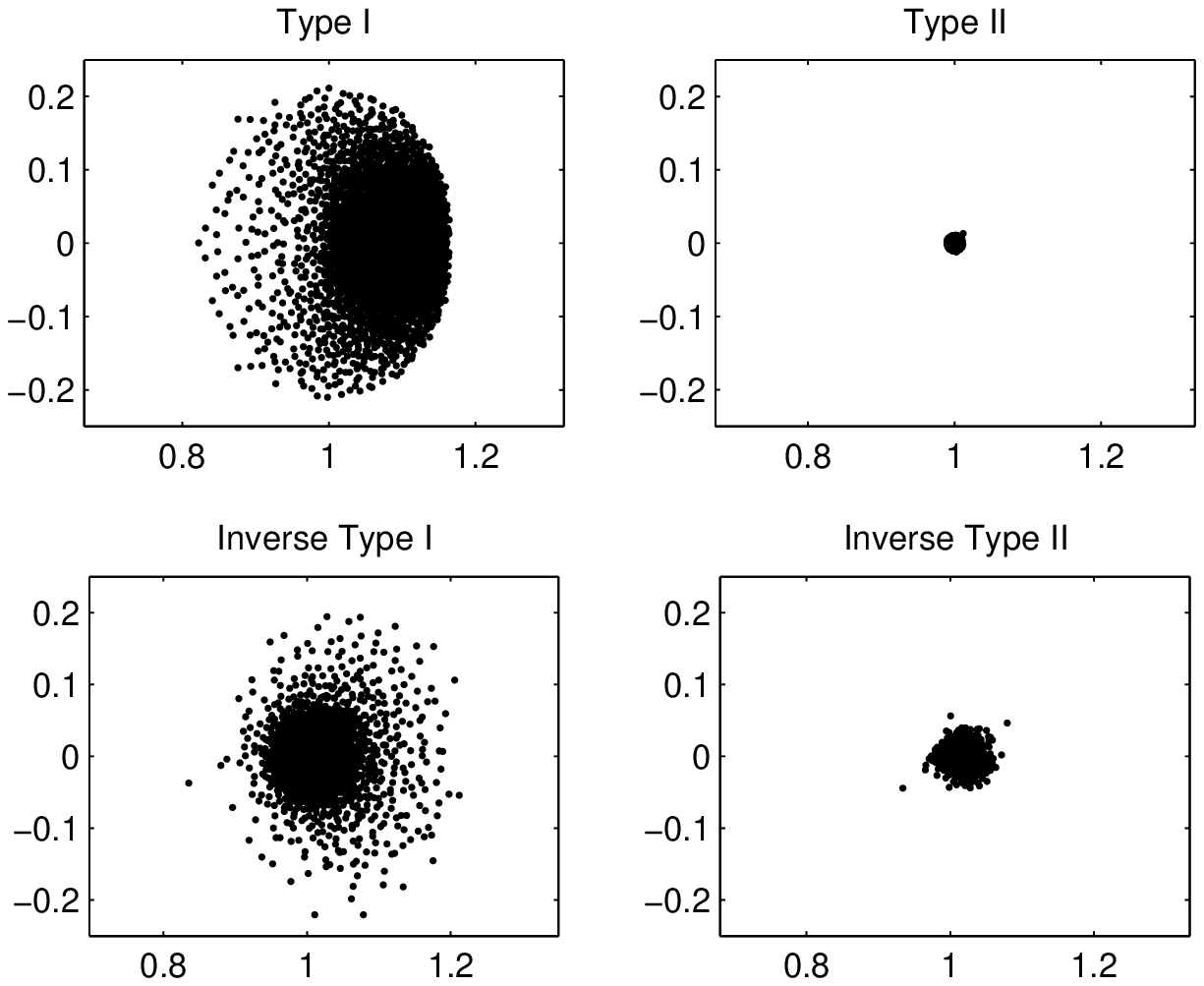}}

\vspace{-0.5cm}
{{\bf Figure 3.}\footnotesize ~~Spectrum of $\left({\tilde S_{bb}^{(2)}}\right)^{-1}S_{bb}$ for various permutations in a SU(3) background at $\beta=5.4$ on a $8^8$ lattice.}

\end{document}